\documentclass[aps,prl,twocolumn,showpacs,byrevtex]{revtex4}

\usepackage{times,mathptm}
\usepackage[dvips]{graphicx,color}
\usepackage{titlesec}
\newcommand{\bqa}{\begin{eqnarray*}}
\newcommand{\eqa}{\end{eqnarray*}}

\begin{document}

\title{Multiband Nature of the Room-Temperature Superconductivity in Compressed LaH$_{10}$}
\author{Chongze Wang, Seho Yi, and Jun-Hyung Cho$^{*}$}
\affiliation{
Department of Physics, Research Institute for Natural Science, and HYU-HPSTAR-CIS High Pressure Research Center, Hanyang
University, 222 Wangsimni-ro, Seongdong-Ku, Seoul 04763, Republic of Korea}
\date{\today}

\begin{abstract}
Recently, the discovery of room-temperature superconductivity (SC) was experimentally realized in the fcc phase of LaH$_{10}$ under megabar pressures. This SC of compressed LaH$_{10}$ has been explained in terms of strong electron-phonon coupling (EPC), but the mechanism of how the large EPC constant and high superconducting transition temperature $T_{\rm c}$ are attained has not yet been clearly identified. Based on the density-functional theory and the Migdal-Eliashberg formalism, we reveal the presence of two nodeless, anisotropic superconducting gaps on the Fermi surface (FS). Here, the small gap is mostly associated with the hybridized states of H $s$ and La $f$ orbitals on the three outer FS sheets, while the large gap arises mainly from the hybridized state of neighboring H $s$ or $p$ orbitals on the one inner FS sheet. Further, we find that the EPC constant of compressed YH$_{10}$ with the same sodalite-like clathrate structure is enhanced due to the two additional FS sheets, leading to a higher $T_{\rm c}$ than LaH$_{10}$. It is thus demonstrated that the multiband pairing of hybridized electronic states is responsible for the large EPC constant and room-temperature SC in compressed hydrides LaH$_{10}$ and YH$_{10}$.
\end{abstract}

\maketitle

The realization of superconductivity (SC) at room temperature is one of the most challenging subjects in modern physics and chemistry. Recently, two experimental groups synthesized the lanthanum hydride LaH$_{10}$ with a sodalite-like clathrate structure [see Fig. 1(a)] at megabar pressures and measured a superconducting transition temperature $T_{\rm c}$ of 250$-$260 K at a pressure of ${\sim}$170 GPa~\cite{ExpLaH10-1,ExpLaH10-2}. Obviously, this record of $T_{\rm c}$ is the highest among so far experimentally available superconducting materials~\cite{H3S, FeH5, CeH9, YH6}, thereby opening a new era of high-$T_{\rm c}$ SC~\cite{Hydride3,Hydride4}.

Historically, a search for room-temperature SC in compressed hydrides dates back to about half a century ago. Based on the Bardeen-Cooper-Schrieffer (BCS) theory~\cite{BCS}, Neil Ashcroft~\cite{Ashc} proposed a pioneering idea that the metallization of hydrogen under high pressures over ${\sim}$400 GPa could exhibit a high-$T_{\rm c}$ SC~\cite{MetalicH1,MetalicH2,MetalicH3}. Since then, in order to achieve metallic hydrogen at relatively lower pressures attainable using diamond anvil cells~\cite{diamondanvil1,diamondanvil2}, many binary hydrides have been theoretically searched~\cite{Hydride3,Hydride4, Wang2012-CaH6, duan2014-H3S, Feng2015-MgH6, Quan2016-H3S, YH10-Boeri, rare-earth-hydride1,rare-earth-hydride2}, among which rare-earth hydrides such as the fcc phases of YH$_{10}$ and LaH$_{10}$ exhibited room-temperature SC at around 200$-$300 GPa~\cite{rare-earth-hydride1,rare-earth-hydride2}. Recently, a number of density functional theory (DFT) studies~\cite{rare-earth-hydride1,rare-earth-hydride2,liangliang-prb2019,fcc-lah10-dft1,fcc-lah10-dft2,chongze2019} have been intensively performed to show that fcc LaH$_{10}$ having a high crystalline symmetry of the space group $Fm$$\overline{3}m$ (No. 225) with the point group O$_h$ features the peculiar bonding characters with anionic La, anionic H$_1$ [forming the squares in Fig. 1(a)], and cationic H$_2$ [forming the hexagons in Fig. 1(a)] atoms, van Hove singularities near the Fermi energy $E_{\rm f}$, and strong electron-phonon coupling (EPC) with H-derived phonon modes. These unique bonding, electronic, and phononic properties of fcc LaH$_{10}$ have been associated with increased EPC constant, leading to the emergence of a room-temperature SC. However, the detailed underlying mechanism of how fcc LaH$_{10}$ forms the large EPC constant and high $T_{\rm c}$ remains to be clarified.

\begin{figure}[ht]
\centering{ \includegraphics[width=8.0cm]{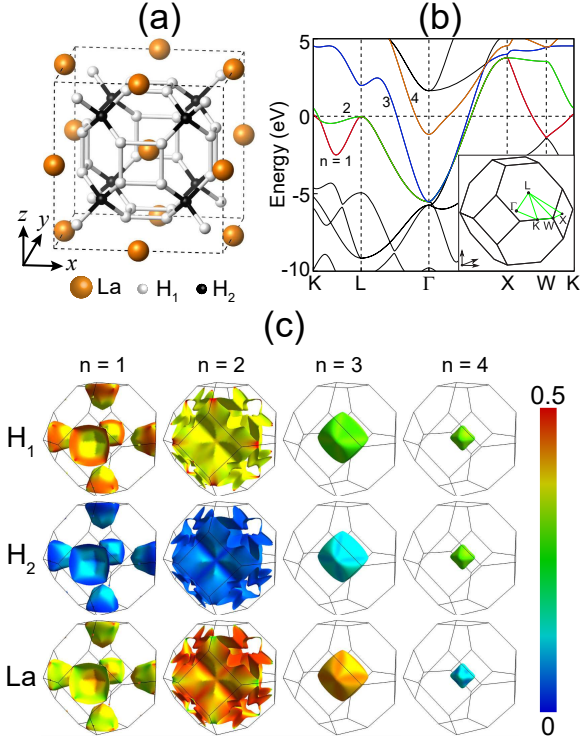} }
\caption{(Color online) (a) Optimized structure of the fcc phase of compressed LaH$_{10}$, composed of the cages of 32 H atoms surrounding a La atom. The two different types of H atoms, i.e., H$_1$ and H$_2$, are drawn with bright and dark circles, respectively. The positive $x$, $y$, and $z$ axes point along the [001], [010], and [001] directions, respectively. (b) Calculated band structure of fcc LaH$_{10}$ and (c) the corresponding FS sheets for the four bands of $n$ = 1, 2, 3, and 4. The inset of (b) shows the Brillouin zone of the fcc primitive cell, where the ${\Gamma}-$X line is parallel to the $x$ axis. In (c), the electronic state at each FS sheet is projected onto H$_1$, H$_2$, and La atoms using the color scale in the range [0, 0.5]. }
\end{figure}

In this Letter, using the DFT calculations~\cite{DFT} and the Migdal-Eliashberg formalism~\cite{Migdal,Eliash,ME-review}, we explore the nature of superconducting gap in fcc LaH$_{10}$. Our analysis of the momentum-resolved superconducting gap reveals that there are two nodeless, anisotropic superconducting gaps on the Fermi surface (FS), representing a two-gap SC with $s$-wave symmetry. It is found that at 20 K, the small gap in the range of 41$-$52 meV is mostly associated with the hybridized states of H$_1$ $s$ and La $f$ orbitals on the three outer FS sheets, while the large gap in the range of 60$-$66 meV arises mainly from the hybridized state of H$_1$ $s$ or $p$ and H$_2$ $s$ orbitals on the one inner FS sheet. Interestingly, for fcc YH$_{10}$ having the same sodalite-like clathrate structure, we find that the four FS sheets whose patterns are similar to those of fcc LaH$_{10}$ slightly shift upward because the two additional FS sheets arising from Y $d$ orbitals are occupied. The resulting six FS sheets of fcc YH$_{10}$ not only contribute to increase the EPC constant due to the increased electron-phonon scattering channels but also produce the two gaps widely distributed in the range of 41$-$75 meV at 20 K, leading to a higher $T_{\rm c}$ than fcc LaH$_{10}$. Therefore, we demonstrate that the underlying mechanism of the large EPC constant and room-temperature SC in compressed hydrides LaH$_{10}$ and YH$_{10}$ can be traced to the multiband pairing of hybridized electronic states.

We first present the electronic band structure of fcc LaH$_{10}$, obtained using the DFT calculations~\cite{methods}. In all the calculations hereafter, we fix a pressure of 300 GPa favoring the fcc phase of LaH$_{10}$~\cite{rare-earth-hydride1, rare-earth-hydride2, liangliang-prb2019, chongze2019}, where the optimized lattice parameters are $a$ = $b$ = $c$ = 4.748 {\AA}. As shown in Fig. 1(b), the calculated band structure exhibits four bands (denoted as $n$ = 1, 2, 3, and 4) crossing $E_{\rm F}$. The corresponding FS sheets are displayed in Fig. 1(c), together with the projection of their electronic states onto the constituent atoms H$_1$, H$_2$, and La. The first FS sheet with the polyhedron shape around the X point is mostly composed of a hybridized state of H$_1$ $s$ and La $f$ orbitals (see Fig. S1 in the Supplemental Material~\cite{supple}). The second FS sheet with the complex shape spreading over the large outer regions of Brillouin zone is also composed of a hybridized state of H$_1$ $s$ and La $f$ orbitals (see Fig. S1 in the Supplemental Material~\cite{supple}). The third and fourth FS sheets are topologically quite similar with concentric polyhedron shapes around the ${\Gamma}$ point. However, these FS sheets have different orbital characters: i.e., the third one arises largely from a hybridized state of H$_1$ $s$ and La $f$ orbitals, whereas the fourth one is mainly due to a hybridized state of H$_1$ $s$ or $p$ and H$_2$ $s$ orbitals (see Fig. S1 in the Supplemental Material~\cite{supple}). It is noticeable that such hybridized electronic states near $E_{\rm F}$ could effectively screen the lattice vibrations involving H atoms, thereby giving rise to a strong EPC in fcc LaH$_{10}$. Further, the FS sheets with different orbital characters are expected to invoke different couplings between the various bands, leading to the emergence of anisotropic multiple SC gaps, as will be demonstrated later.

To illustrate an overview of the EPC and $T_{\rm c}$ of fcc LaH$_{10}$, we calculate the Eliashberg function ${\alpha}^{2}F({\omega})$ and integrated EPC constant ${\lambda}({\omega})$ using the isotropic Migdal-Eliashberg equations~\cite{Migdal,Eliash,ME-review}. The calculated results together with the logarithmically average phonon frequency ${\omega}_{\rm log}$ are displayed in Fig. 2(a). We find that ${\lambda}({\omega})$ increases monotonously as ${\omega}$ increases up to the optical phonon modes with a high frequency of ${\sim}$1800 cm$^{-1}$, indicating that the lattice vibrational modes in the whole frequency range participate in the increase of ${\lambda}({\omega})$. It is noted that the four FS sheets distributed over the whole Brillouin zone provide the EPC channels with many phonon modes of widely distributed ${\bf q}$ wavevectors. Using the McMillan-Allen-Dynes formula~\cite{Allen}, we estimate $T_{\rm c}$ as a function of ${\omega}$ [see Fig. 2(a)]. Although the McMillan-Allen-Dynes formula cannot properly describe anisotropic multiband SC, its estimation of $T_{\rm c}({\omega})$ may give a qualitative aspect of how largely certain-frequency phonon modes contribute to $T_{\rm c}$. As shown in Fig. 2(a), H-derived optical phonon modes contribute to a nearly linear increase of $T_{\rm c}$ with increasing ${\omega}$. It is, however, interesting to note that, even though the acoustic phonon modes of La atoms provide ${\sim}$12\% of the total EPC constant ${\lambda}$ = ${\lambda}$(${\infty}$)~\cite{liangliang-prb2019, chongze2019}, they hardly contribute to an increase of $T_{\rm c}$ [see Fig. 2(a)].

\begin{figure}[ht]
\centering{ \includegraphics[width=8.0cm]{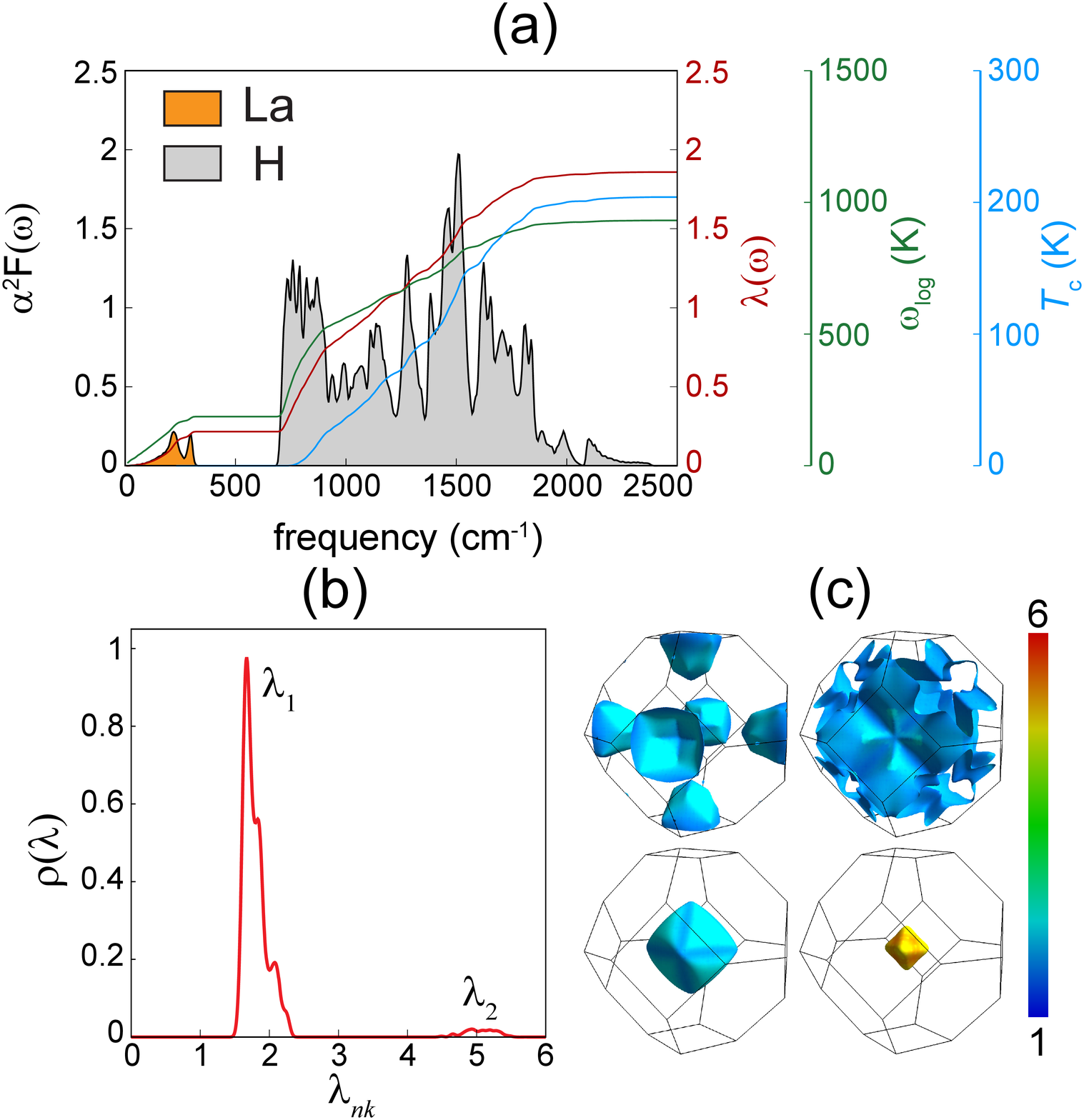} }
\caption{(Color online) (a) Isotropic Eliashberg function ${\alpha}^{2}F({\omega})$ (black), integrated EPC constant ${\lambda}({\omega})$ (red), ${\omega}_{\rm log}$ (green), and $T_{\rm c}$ (blue) of fcc LaH$_{10}$. (b) Distribution of ${\bf k}$-resolved EPC constant ${\lambda}_{n{\bf k}}$ and (c) the projected ${\lambda}_{n{\bf k}}$ on each FS sheet. In (b), the two separated regimes of ${\lambda}_{n{\bf k}}$ are indicated as ${\lambda}_1$ and ${\lambda}_2$.}
\end{figure}

In order to examine the anisotropy in the EPC of fcc LaH$_{10}$, we use the anisotropic Migdal-Eliashberg equations~\cite{Migdal,Eliash,ME-review} to calculate the ${\bf k}$-resolved EPC constant ${\lambda}_{n{\bf k}}$ for the electronic state ($n$, ${\bf k}$), which includes all available electron-phonon scattering processes connecting ${\bf k}$ and other ${\bf k}$ points on the FS sheets. The distribution of ${\lambda}_{n{\bf k}}$ and their projection on each FS sheet are plotted in Figs. 2(b) and 2(c), respectively. We find that there are two regimes of ${\lambda}_{n{\bf k}}$: i.e., the lower regime ${\lambda}_1$ in the range of 1.55$-$2.29 arising from the $n$ = (1, 2, 3) bands and the upper one ${\lambda}_2$ in the range of 4.49$-$5.49 arising from the $n$ = 4 band. Here, it is noticeable that (i) the larger distribution of ${\lambda}_1$ is caused by the high density of states (DOS) of the hybridized $n$ = (1, 2, 3) states at $E_{\rm F}$, and (ii) the larger ${\lambda}_2$ values represent the strong EPC of the hybridized $n$ = 4 state  stemming from H$_1$ $s$ or $p$ and H$_2$ $s$ orbitals. Thus, the division of the two regimes ${\lambda}_1$ and ${\lambda}_2$ indicates the different EPC strengths depending on the orbital characters of the hybridized electronic states on the FS sheets. We also note that some spreads of the ${\lambda}_1$ and ${\lambda}_2$ values feature the anisotropy of EPC, which is apparent from the projection of ${\lambda}_{n{\bf k}}$ on each FS sheet: i.e., the size of ${\lambda}_{n{\bf k}}$ changes with respect to the ${\bf k}$ directions on each FS sheet [see Fig. 2(c)].

It is natural that the two well-separated regimes of ${\lambda}_{n{\bf k}}$ in fcc LaH$_{10}$ could give rise to two superconducting gaps. By numerically solving the anisotropic Migdal-Eliashberg equations~\cite{Migdal,Eliash,ME-review} with a typical Coulomb pseudopotential parameter of ${\mu}^*$ = 0.13~\cite{rare-earth-hydride1,rare-earth-hydride2}, we calculate the temperature dependence of superconducting gap ${\Delta}$. Figure 3(a) displays the energy distribution of ${\Delta}$ as a function of temperature. We find that there are the two gaps ${\Delta}_1$ and ${\Delta}_2$, indicating a two-gap SC. These two gaps close at $T_{\rm c}$ ${\approx}$ 252 K [see Fig. S2(a) in the Supplemental Material~\cite{supple}]. For $T$ $<$ 100 K, ${\Delta}_1$ is distributed in the range of 41$-$52 meV, while ${\Delta}_2$ in the range of 60$-$66 meV. From the ${\bf k}$-resolved superconducting gap ${\Delta}_{n{\bf k}}$ on the four FS sheets [see Fig. 3(b)], we find that the formation of ${\Delta}_1$ is associated with the FS sheets of $n$ = 1, 2, and 3, while that of ${\Delta}_2$ arises from the FS sheet of $n$ = 4. Thus, we can say that ${\lambda}_{n{\bf k}}$ and ${\Delta}_{n{\bf k}}$ are correlated with each other: i.e., the larger the magnitude of ${\lambda}_{n{\bf k}}$, the higher is the ${\Delta}_{n{\bf k}}$ value. It is noticeable that the size of ${\Delta}_{n{\bf k}}$ changes on each FS sheet without any node, representing the anisotropic superconducting gaps with s-wave symmetry. In Fig. 3(a), the ${\Delta}$ values obtained from the isotropic Migdal-Eliashberg formalism are also plotted with the dashed line. Here, the gap closes around $T_{\rm c}$ ${\approx}$ 233 K, smaller than that ($T_{\rm c}$ ${\approx}$ 252 K) estimated from the anisotropic Migdal-Eliashberg formalism, but both values are much larger than that (${\sim}$204 K) obtained using the McMillan-Allen-Dynes formula~\cite{Allen}.

\begin{figure}[ht]
\centering{ \includegraphics[width=8.0cm]{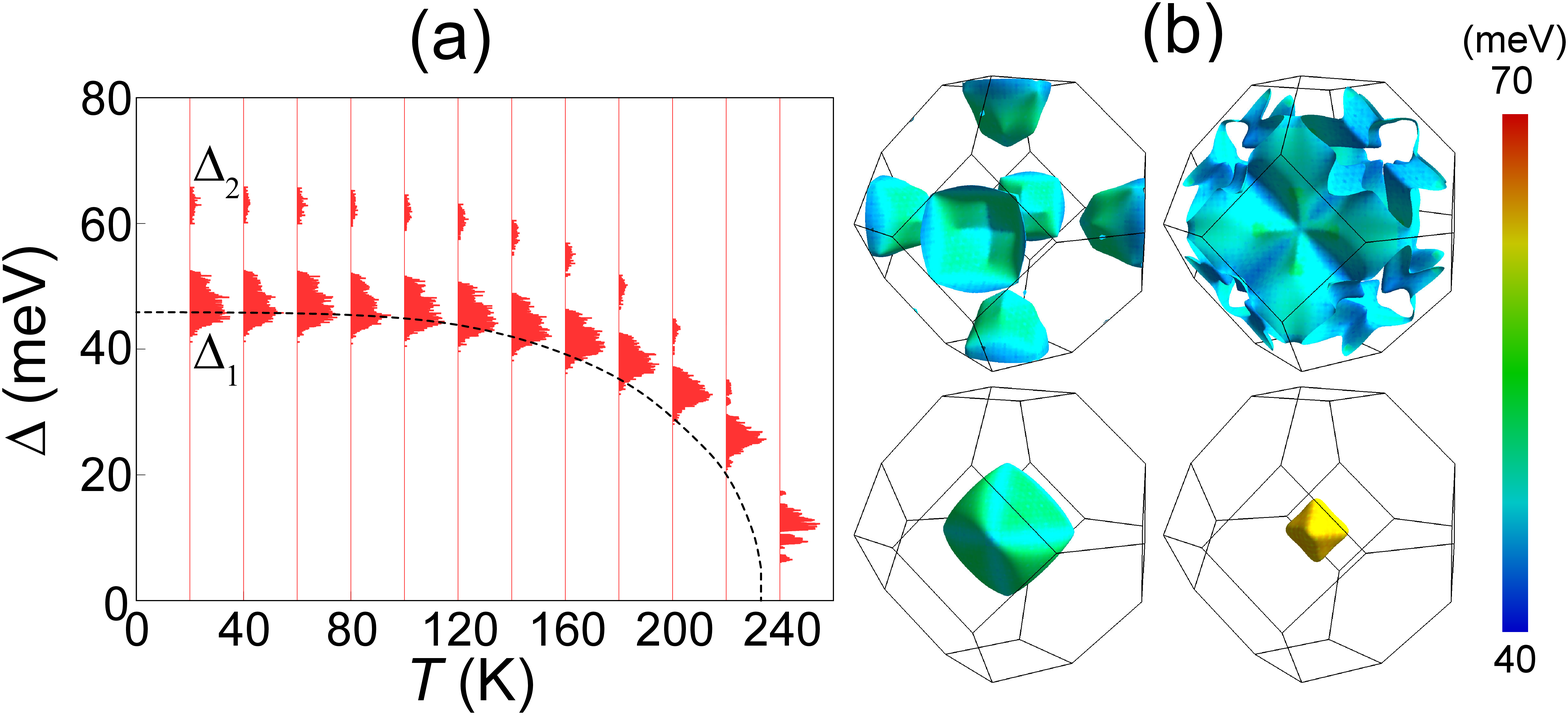} }
\caption{(Color online) (a) Energy distribution of the anisotropic superconducting gap ${\Delta}$ of fcc LaH$_{10}$. The two separated gaps are indicated as ${\Delta}_1$ and ${\Delta}_2$. The dashed line in (a) represents the ${\Delta}$ values, estimated using the isotropic Migdal-Eliashberg equations. (b) ${\bf k}$-resolved superconducting gap ${\Delta}_{n{\bf k}}$ on the four FS sheets, computed at 20 K. }
\end{figure}

\begin{figure}[htb]
\centering{ \includegraphics[width=8.0cm]{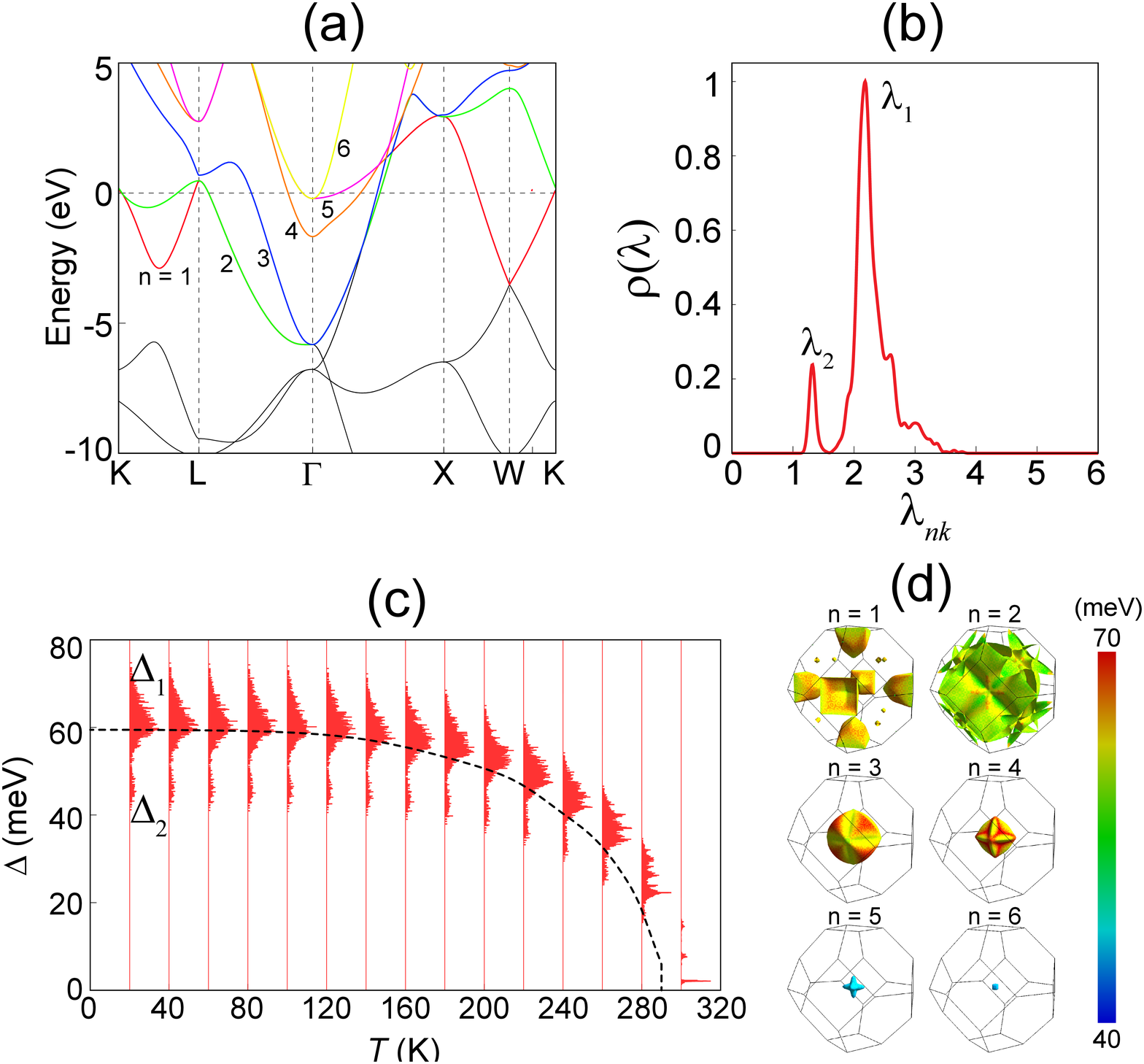} }
\caption{(Color online) (a) Calculated band structure of fcc YH$_{10}$ at 300 GPa and (b) the distribution of ${\lambda}_{n{\bf k}}$. In (b), the two regimes of ${\lambda}_{n{\bf k}}$ are indicated as ${\lambda}_1$ and ${\lambda}_2$. (c) Energy distribution of ${\Delta}$ as a function of temperature and (d) ${\Delta}_{n{\bf k}}$ on the six FS sheets, computed at 20 K. The dashed line in (c) represents ${\Delta}$ estimated using the isotropic Migdal-Eliashberg equations.}
\end{figure}

Based on our results of ${\lambda}_{n{\bf k}}$ and ${\Delta}_{n{\bf k}}$ in fcc LaH$_{10}$, we conclude that the predicted anisotropic EPC and superconducting gaps are attributed to the multiband pairing with the four FS sheets composed of the hybridized states of different orbital characters [see Fig. 1(c)]. To further demonstrate the importance of multiband pairing in high-$T_{\rm c}$ SC, we study the properties of ${\lambda}_{n{\bf k}}$ and ${\Delta}_{n{\bf k}}$ of fcc YH$_{10}$ that was previously~\cite{rare-earth-hydride1,rare-earth-hydride2,YH10-Boeri} predicted to exhibit higher $T_{\rm c}$ than fcc LaH$_{10}$. The geometry of fcc YH$_{10}$ is the same sodalite-like clathrate structure as fcc LaH$_{10}$ (see Fig. S3 in the Supplemental Material~\cite{supple}). Figure 4(a) shows the calculated electronic band structure of fcc YH$_{10}$ at 300 GPa. We find that fcc YH$_{10}$ has very similar dispersions for the four bands of $n$ = 1, 2, 3, and 4, as compared with those of fcc LaH$_{10}$ [see Fig. 1(b)]. However, in fcc YH$_{10}$, the two electron-like bands of $n$ = 5 and 6 originating mostly from Y $d$ orbitals become occupied below $E_{\rm F}$ around the ${\Gamma}$ point [see Fig. 4(a) and Fig. S4 in the Supplemental Material~\cite{supple}]. Therefore, fcc YH$_{10}$ has the six FS sheets which provide more EPC channels, leading to the enhancements of the EPC constant and $T_{\rm c}$. Using the isotropic Migdal-Eliashberg formalism~\cite{Migdal,Eliash,ME-review}, we obtain ${\lambda}$ = 2.23 and $T_{\rm c}$ = 290 K, larger than those (${\lambda}$ = 1.86 and $T_{\rm c}$ = 233 K) of fcc LaH$_{10}$. Figure 4(b) shows the calculated distribution of anisotropic ${\lambda}_{n{\bf k}}$ in fcc YH$_{10}$, where the peaks of the two regimes ${\lambda}_1$ and ${\lambda}_2$ locate at around 2.19 and 1.33, respectively. From the projection of ${\lambda}_{n{\bf k}}$ on each FS sheet, we find that ${\lambda}_1$ originates mostly from the $n$ = (1, 2, 3, 4) bands, while ${\lambda}_2$ is due to the $n$ = (5, 6) bands [see Fig. S5 in the Supplemental Material~\cite{supple}]. Compared with the large ${\lambda}_{n{\bf k}}$ values (between 4.49$-$5.49) originating from the $n$ = 4 band of fcc LaH$_{10}$ [see Figs. 2(b) and 2(c)], the corresponding values of fcc YH$_{10}$ are much reduced to about 2.20$-$3.79 (see Fig. S5), which reflects the weakened intraband and interband coupling strengths of the $n$ = 4 band as the $n$ = 5 and 6 bands are occupied. However, the ${\lambda}_{n{\bf k}}$ values of the $n$ = (1, 2, 3) bands in fcc YH$_{10}$ increase by ${\sim}$0.5, compared to the corresponding values in fcc LaH$_{10}$ [see the peak positions of ${\lambda}_1$ in Figs. 2(b) and 4(b), as well as the the projected ${\lambda}_{n{\bf k}}$ on each FS sheet in Figs. 2(c) and S5]. It is noted that the $n$ = (5, 6) bands having weak hybridization characters (see Fig. S4 in the Supplemental Material~\cite{supple}) give rise to  relatively smaller values of ${\lambda}_2$ [see Fig. 4(b)]. Figures 4(c) and 4(d) show the temperature dependence of ${\Delta}$ and the ${\bf k}$-resolved ${\Delta}_{n{\bf k}}$ on the six FS sheets, respectively. These results agree well with those obtained by a recent first-principles calculation of Heil $et$ $al$.~\cite{YH10-Boeri}. As shown in Fig. 4(c), for $T$ $<$ 100 K, ${\Delta}$ consisting of the two gaps ${\Delta}_1$ and ${\Delta}_2$ is widely distributed between 41 and 75 meV. This broad gap closes at $T_{\rm c}$ ${\approx}$ 308 K [see Fig. S2(b) in the Supplemental Material~\cite{supple}], higher than that [${\sim}$252 K in Fig. 3(a)] of fcc LaH$_{10}$. We note that the large gap ${\Delta}_1$ arises from the $n$ = (1, 2, 3, 4) bands, while the small gap ${\Delta}_2$ is due to the $n$ = (5, 6) bands [see Fig. 4(d)]. It is thus demonstrated that the multiband pairing of fcc YH$_{10}$ with the six FS sheets gives rise to higher $T_{\rm c}$ than fcc LaH$_{10}$ with the four FS sheets.

Since the minima of the $n$ = (5, 6) bands of fcc YH$_{10}$ are close to $E_{\rm F}$, they are enabled to be unoccupied by the hole doping of $n_h$ $>$ ${\sim}$0.17$e$ per unit cell (see Fig. S6 in the Supplemental Material~\cite{supple}). In order to examine how the occupation of the $n$ = (5, 6) bands influences SC, we use the isotropic Migdal-Eliashberg formalism~\cite{Migdal,Eliash,ME-review} to estimate the variations of ${\lambda}$ and $T_{\rm c}$ with respect to the amount of $n_h$.
We find that ${\lambda}$ and $T_{\rm c}$ decrease monotonously with increasing $n_h$ (see Table SI in the Supplemental Material~\cite{supple}). As a result, $T_{\rm c}$ decreases from 290 K (without hole doping) to 269 K at $n_h$ = 0.3$e$. It is thus manifested that the reduced number of FS sheets via hole doping decreases the EPC channels, resulting in a decrease of $T_{\rm c}$. Interestingly, the hole doping in fcc LaH$_{10}$ slightly increases $T_{\rm c}$ from 233 K (without hole doping) to 245 K at $n_h$ = 0.3$e$ (see Table SI). We note that for $n_h$ = 0.3$e$, the predicted $T_{\rm c}$ value of fcc LaH$_{10}$ is still lower than that of fcc YH$_{10}$ by 24 K. Considering that both the hole-doped systems with $n_h$ = 0.3$e$ have the same number of FS sheets, it is implied that fcc YH$_{10}$ would have larger electron-phonon matrix elements than fcc LaH$_{10}$.

In conclusion, our first-principles calculations for fcc LaH$_{10}$ have shown that the hybridized states of La and H$_1$ atoms as well as H$_1$ and H$_2$ atoms on the four FS sheets are strongly coupled with the phonon modes in the whole frequency range, contrasting with a typical low-T$_{\rm c}$ BCS-type superconductor MgB$_2$~\cite{MgB2} where certain phonon modes significantly contribute to the EPC. Specifically, we reveled that the presence of such multiple FS sheets with different orbital characters gives rise to the two nodeless, anisotropic superconducting gaps on the FS. It is thus demonstrated that the two factors such as multiband pairing and the hybridized states of constituent atoms play crucial roles in determining the recently observed~\cite{ExpLaH10-1,ExpLaH10-2} room-temperature SC in fcc LaH$_{10}$. The present findings have important implications for understanding the underlying mechanisms involved in high-$T_{\rm c}$ SC of compressed hydrides, as well as in tuning $T_{\rm c}$ through carrier doping.

\noindent {\bf Acknowledgement.}
This work was supported by the National Research Foundation of Korea (NRF) grant funded by the Korean Government (Grants No. 2019R1A2C1002975, No. 2016K1A4A3914691, and No. 2015M3D1A1070609). The calculations were performed by the KISTI Supercomputing Center through the Strategic Support Program (Program No. KSC-2018-CRE-0063) for the supercomputing application research.  \\

\noindent $^{*}$ Corresponding author: chojh@hanyang.ac.kr


\newpage

\onecolumngrid
\newpage
\titleformat*{\section}{\LARGE\bfseries}

\renewcommand{\thefigure}{S\arabic{figure}}
\setcounter{figure}{0}

\vspace{1.2cm}

\section{Supplemental Material for Multiband Nature of the Room-Temperature Superconductivity in Compressed LaH$_{10}$}
\vspace{1.2cm}
\begin{flushleft}

{\bf 1. Band projections onto the orbitals of La, H$_1$, and H$_2$ atoms in fcc LaH$_{10}$.}
\begin{figure}[ht]
\includegraphics[width=16cm]{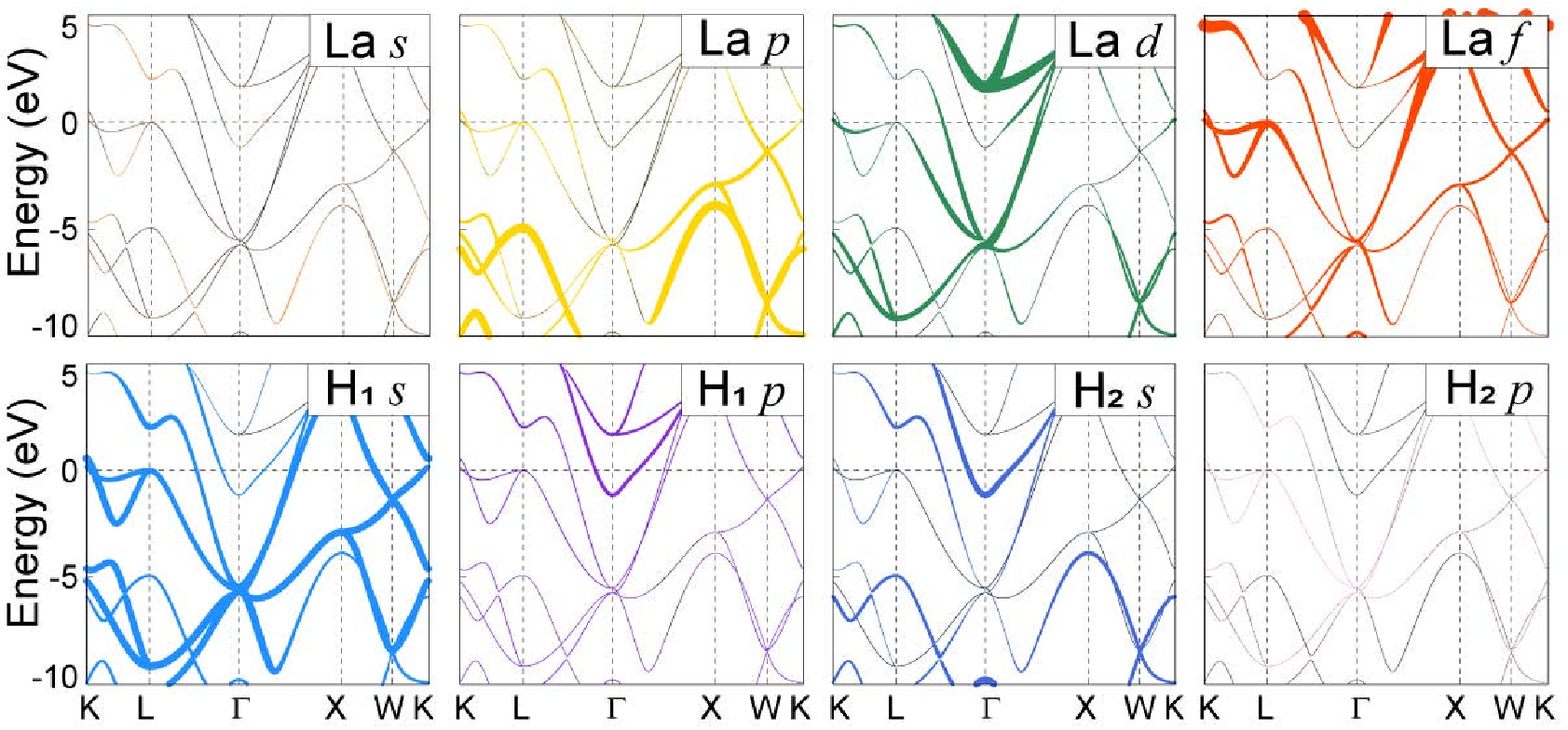}
\caption{ Calculated bands projected onto the La $s$, La $p$, La $d$, La $f$, H$_1$ $s$, H$_1$ $p$, H$_2$ $s$, and H$_2$ $p$ orbitals in fcc LaH$_{10}$. Here, the radii of circles are proportional to the weights of the corresponding orbitals. }
\end{figure}

\vspace{1.2cm}

{\bf 2. Estimation of $T_{\rm c}$ using the anisotropic Migdal-Eliashberg equations.}
\begin{figure}[ht]
\includegraphics[width=12cm]{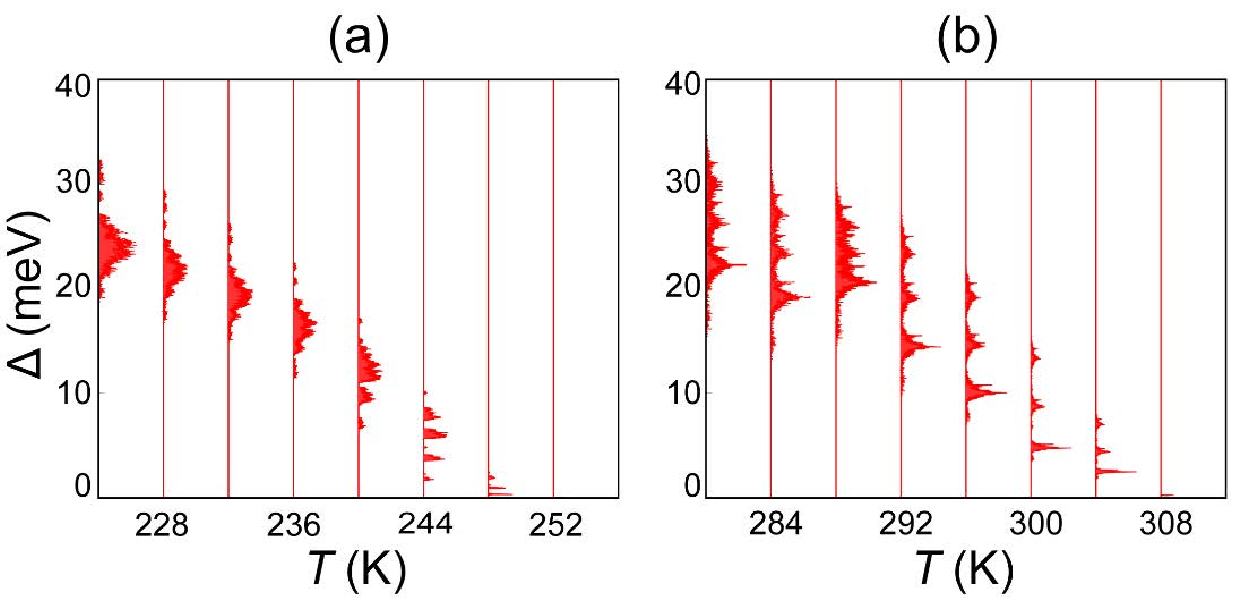}
\caption{ Calculated anisotropic superconducting gap $\Delta$ of (a) fcc LaH$_{10}$ and (b) fcc YH$_{10}$ near $T_{\rm c}$.}
\end{figure}

\newpage

{\bf 3. Optimized structure of the fcc phase of compressed YH10.}
\begin{figure}[ht]
\includegraphics[width=8cm]{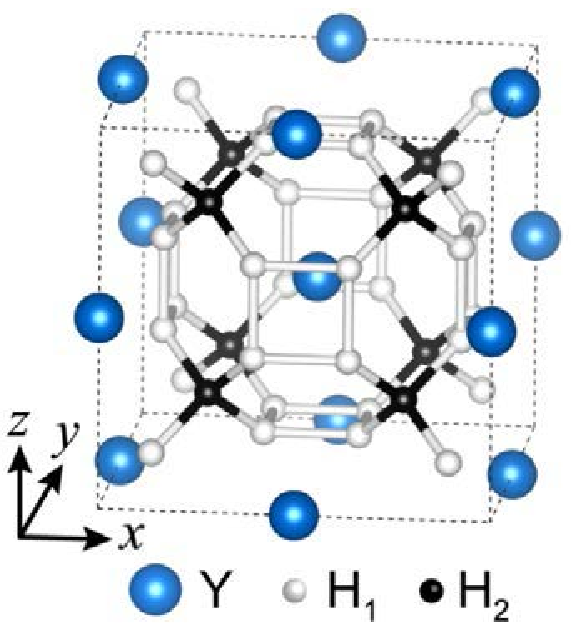}
\caption{ Optimized structure of the fcc phase of compressed YH$_{10}$, composed of the cages of 32 H atoms surrounding a Y atom. The two different types of H atoms, i.e., H$_1$ and H$_2$, are drawn with bright and dark circles, respectively. The positive $x$, $y$, and $z$ axes point along the [001], [010], and [001] directions, respectively. The lattice parameters are $a$ = $b$ = $c$ = 4.598 \AA at 300 GPa.}
\end{figure}

\vspace{1.2cm}

{\bf 4. Band projections onto the orbitals of Y, H$_1$, and H$_2$ atoms in fcc YH$_{10}$.}
\begin{figure}[ht]
\includegraphics[width=16cm]{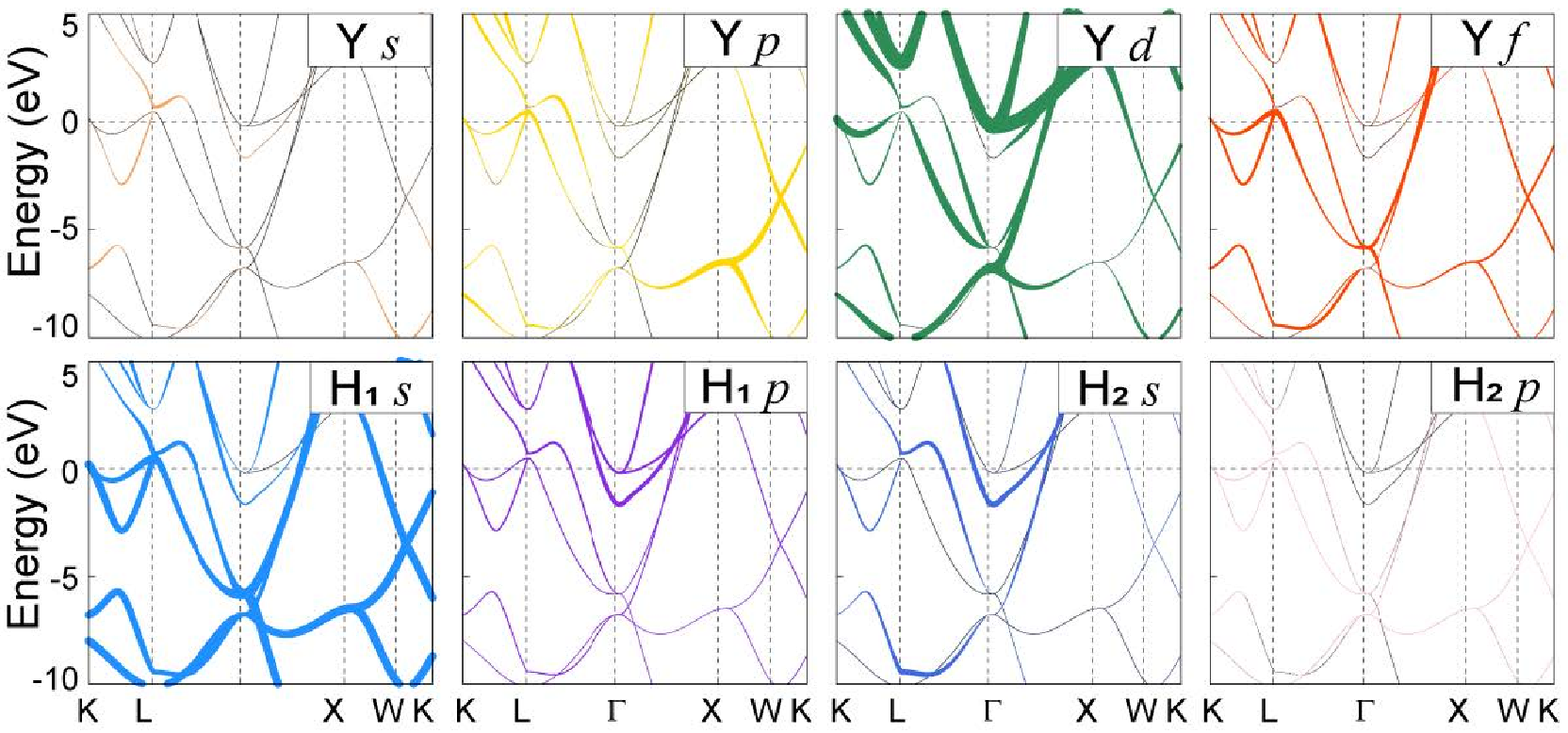}
\caption{Calculated bands projected onto the Y $s$, Y $p$, Y $d$, Y $f$, H$_1$ $s$, H$_1$ $p$, H$_2$ $s$, and H$_2$ $p$ orbitals in fcc YH$_{10}$. Here, the radii of circles are proportional to the weights of the corresponding orbitals. }
\end{figure}

\newpage

{\bf 5. Projection of ${\lambda}_{n{\bf k}}$ on each FS sheet of fcc YH$_{10}$.}
\begin{figure}[ht]
\includegraphics[width=12cm]{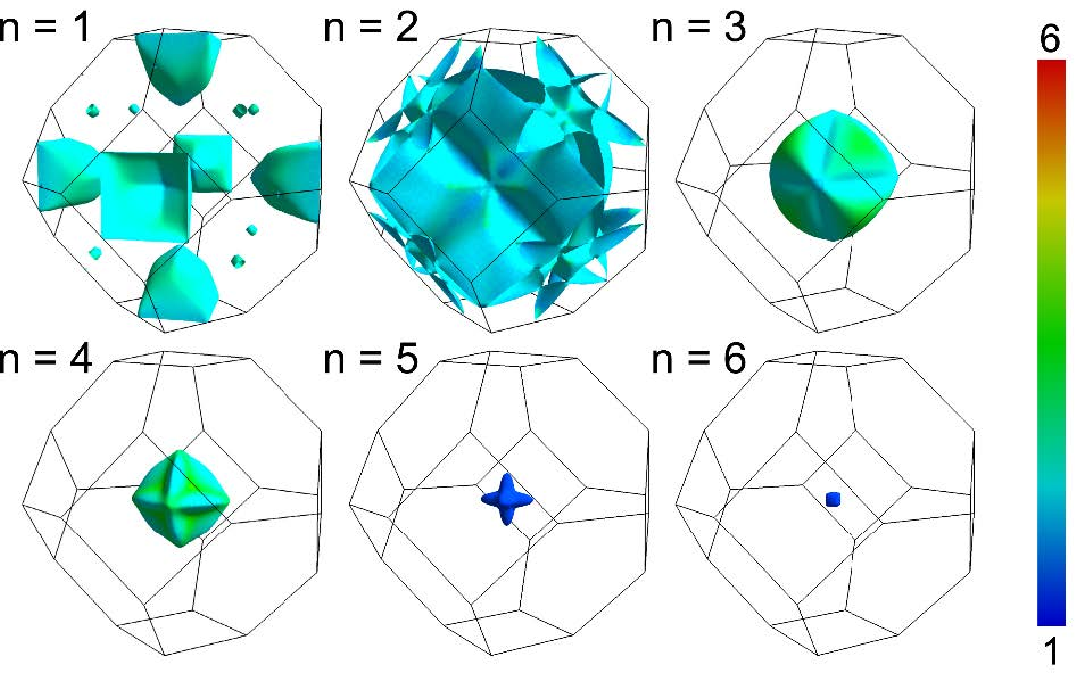}
\caption{ Projection of ${\lambda}_{n{\bf k}}$ on the six Fermi surface sheets of fcc YH10.}
\end{figure}

\vspace{1.2cm}

{\bf 6. Band structure of fcc YH$_{10}$ with hole doping.}
\begin{figure}[ht]
\includegraphics[width=12cm]{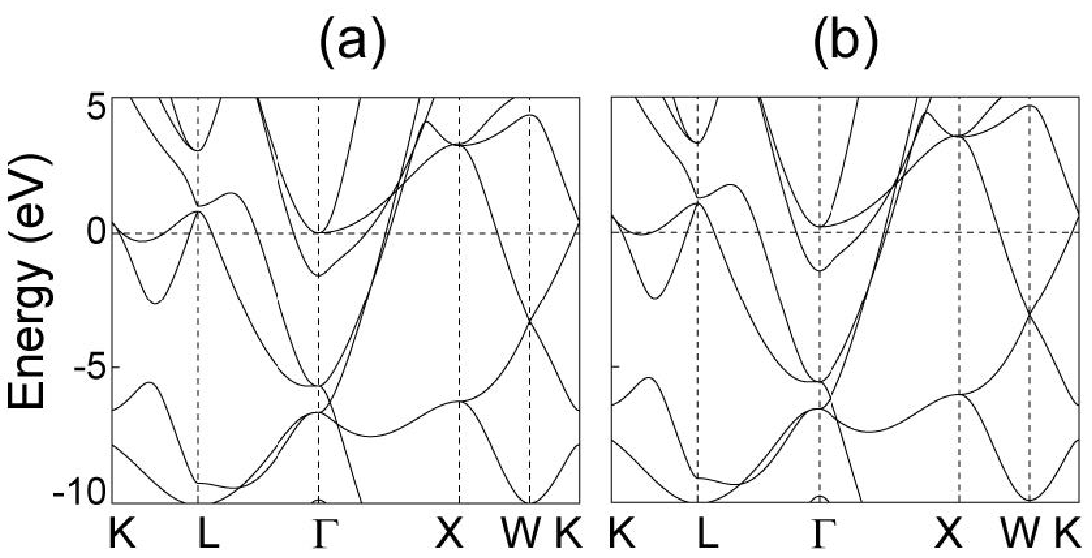}
\caption{ Calculated band structures of fcc YH$_{10}$ with the amount of hole doping $n_h$ = (a) 0.17$e$ and (b) 0.3$e$ per unit cell.}
\end{figure}

\newpage

{\bf 7. Band structures of fcc LaH$_{10}$ and fcc YH$_{10}$, obtained using PAW+PBE pseudopo-tentials in VASP and ONCV+PBE pseudopotentials in QE.}
\begin{figure}[ht]
\includegraphics[width=12cm]{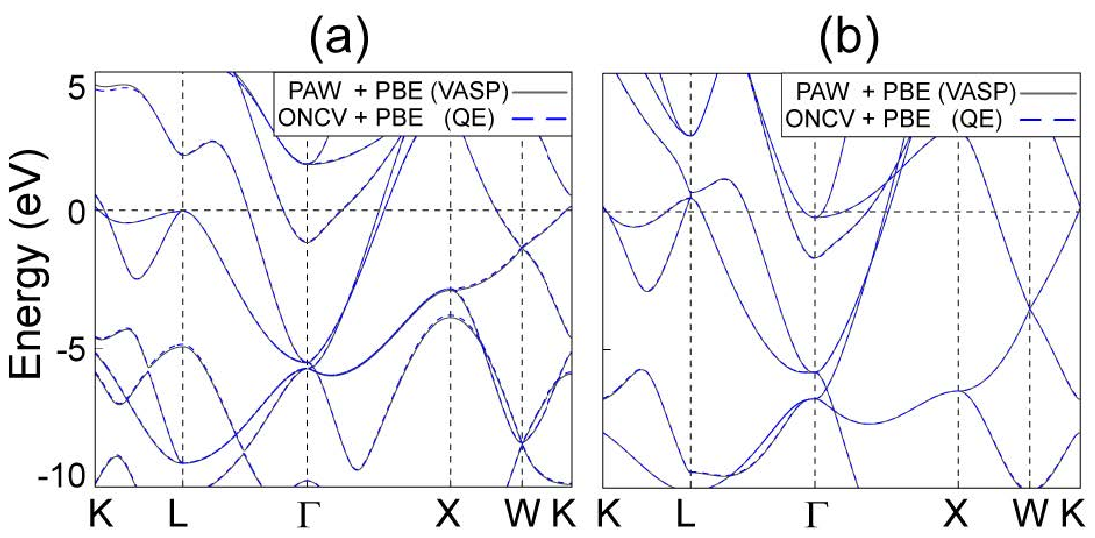}
\caption{ Calculated electronic band structures of (a) fcc LaH$_{10}$ and (b) fcc YH$_{10}$ at 300 GPa. The two results obtained using PAW+PBE pseudopotentials in VASP (grey lines) and ONCV+PBE pseudopotentials in QE (blue dashed lines) overlap for comparison. We note that (i) the lattice constants optimized using PAW+PBE pseudopotentials in VASP and ONCV+PBE pseudopotentials in QE change only by less than 0.1\%, and (ii) the band structures obtained using the two different pseudopotentials are nearly the same with each other, especially near the Fermi energy.}
\end{figure}

\vspace{1.2cm}

{\bf Table SI. Calculated $\lambda$ and $T_{\rm c}$ values of fcc YH$_{10}$ as a function of hole doping.}
\begin{table}[ht]
\centering
\caption{Calculated ${\lambda}$ and $T_{\rm c}$ values of fcc YH$_{10}$ at $n_h$ = 0.1, 0.2, and 0.3$e$ per unit cell using the isotropic Migdal-Eliashberg formalism with ${\mu}^*$ = 0.13. For comparison, the corresponding values of fcc LaH$_{10}$ are given in parentheses. The results without hole doping are also given in the line of $n_h$ = 0. }
\begin{ruledtabular}
\begin{tabular}{ccccr}
$n_h$ ($e$/unit cell) & ${\lambda}$ &$T_{\rm c}$ (K)\\
\hline
0   &  2.23 (1.86)   & 290 (233) \\
0.1   &  2.11 (1.89)   & 282 (235) \\
0.2   &  2.00 (1.97)   & 275 (239) \\
0.3   &  1.91 (2.11)    & 269 (245)
\end{tabular}
\end{ruledtabular}
\end{table}

\end{flushleft}
\end{document}